\documentclass[pdflatex,sn-mathphys]{sn-jnl}

\jyear{2021}%

\theoremstyle{thmstyleone}%
\theoremstyle{thmstyletwo}%

\theoremstyle{thmstylethree}%

\usepackage{siunitx}

\begin{document}

\title{Two-sided Acoustic Metascreen for Broadband and Individual Reflection and Transmission Control}


\author[1,2]{\fnm{Ao} \sur{Chen}}\email{aochen@bu.edu}

\author*[1,2]{\fnm{Xin} \sur{Zhang}}\email{xinz@bu.edu}

\affil[1]{\orgdiv{Department of Mechanical Engineering}, \orgname{Boston University}, \orgaddress{\city{Boston}, \postcode{02215}, \state{MA}, \country{USA}}}

\affil[2]{\orgdiv{Photonics Center}, \orgname{Boston University}, \orgaddress{\city{Boston}, \postcode{02215}, \state{MA}, \country{USA}}}


\abstract{Acoustic wave modulation plays a pivotal role in various applications, including sound-field reconstruction, wireless communication, and particle manipulation, among others. However, current acoustic metamaterial and metasurface designs typically focus on controlling either reflection or transmission waves, often overlooking the coupling between amplitude and phase of acoustic waves. To fulfill this gap, we propose and experimentally validate a design enabling complete control of reflected and transmitted acoustic waves individually across a frequency range of 4 kHz to 8 kHz, allowing arbitrary combinations of amplitude and phase for reflected and transmitted sound in a broadband manner. Additionally, we demonstrate the significance of our approach for sound manipulation by achieving acoustic diffusion, reflection, focusing, and generating a two-sided 3D hologram at three distinct frequencies. These findings open an alternative avenue for extensively engineering sound waves, promising applications in acoustics and related fields.}

\maketitle
\newpage
\section{Introduction}\label{sec1}
The pursuit of wave manipulation, encompassing electromagnetic and acoustic waves, has long captivated scientific inquiry. Metamaterials have emerged as a pivotal area of study\cite{cummer2016controlling,assouar2018acoustic,zheludev2012metamaterials}, demonstrating remarkable advancements primarily due to their capacity to exert precise control over electromagnetic waves\cite{zheludev2012metamaterials,sun2012gradient,yu2011light,ni2012broadband,cui2014coding,chen2020directional,liu2014broadband}. This capability stems from their subwavelength artificial structures, allowing for the manipulation of their effective permittivity ($\varepsilon$) and permeability ($\mu$) in arbitrary manners. This versatility has unveiled promising applications, including anomalous refraction and reflection
\cite{pendry2000negative,shelby2001experimental,shalaev2005negative,silveirinha2006tunneling}, the development of metalenses\cite{fang2005sub,wang2021high}, and the conversion of vortex beams\cite{grier2003revolution,padgett2011tweezers}. In the realm of acoustics\cite{cummer2016controlling,assouar2018acoustic,fleury2015invisible,li2018systematic,melde2016holograms,fu2022asymmetric,fang2006ultrasonic,ozcelik2018acoustic}, metamaterials composed of locally resonant unit cells have demonstrated the capacity to govern sound throughout its emission\cite{ji2024metamaterial,dong2020bioinspired}, propagation\cite{wu2022metamaterial,gao2021emitting,chen2022broadband,chen2023composite}, and reception\cite{zhang2021remote,ji2023metamaterial}. Furthermore, analogous to their electromagnetic counterparts, two-dimensional appearances of metamaterials, known as acoustic metasurfaces\cite{assouar2018acoustic,ma2014acoustic,tian2019programmable,ghaffarivardavagh2018horn,tang2018hollow}, facilitate the control of wavefronts by implementing local phase modulation, adhering to principles akin to the generalized Snell's law\cite{yu2011light,ni2012broadband}. The advent of metasurfaces heralds novel avenues for various applications, such as holography\cite{zhu2021systematic,xu2023sound,li2023ultrathin}, invisibility cloaking\cite{faure2016experiments,esfahlani2016acoustic}, and precise focusing of waves\cite{chen2024angle}. \\
Despite the fascinating wave-steering effects, conventional metasurfaces are primarily reflection-based. This is because a high reflectivity can be easily ensured by integrating a rigid boundary into the structure. Therefore, the development of transmission-type metasurfaces, which enable high transmission while also facilitating precise phase control simultaneously, presents a greater challenge. Acoustically, the incorporation of resonance-based unit cells such as Helmholtz Resonators and space-coiling metamaterials has demonstrated successful phase modulations on the transmission side of wave propagation\cite{tian2019programmable,fang2006ultrasonic,zhu2021systematic}. However, it is important to recognize that conventional metamaterials and metasurfaces typically operate exclusively in either reflection-type or transmission-type modes, with relatively limited exploration in the domain where devices can effectively function on both sides simultaneously. \\
Recently, electromagnetic metamaterials have showcased the ability to exert full-space control over both transmitted and reflected wavefronts by manipulating wave polarization\cite{shang2021transmission,zhang2018transmission,wang2021reconfigurable}. However, due to the distinct polarization characteristics of acoustic waves in comparison to electromagnetic waves, achieving two-sided wavefront modulations in the field of acoustics has been rare\cite{fleury2014sound,zhu2021janus,ding2019experimental,tang2022transmission,shi2022broadband}. While earlier approaches relied on external biases (e.g., 
airflow)\cite{fleury2014sound,zhu2021janus,ding2019experimental}, termed Janus metamaterials, to achieve wave control on both sides of the interface, these methods demonstrated effective functionality within a relatively narrow frequency range. Furthermore, the intricate geometries and confined channels inherent in these approaches have posed significant challenges in the fabrication process and have resulted in considerable viscous energy losses, particularly as the dimensions decrease. Therefore, there is a pressing need for the development of a design capable of broadband control over both reflected and transmitted acoustic waves, without the reliance on complex and cumbersome structures. 
\begin{figure}[!h]
\centering
\includegraphics[width=1\textwidth]{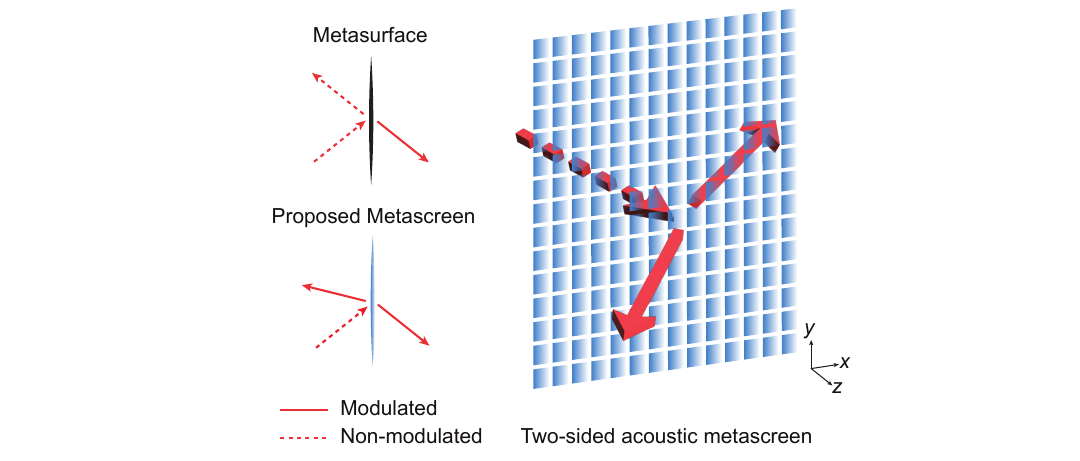}
\caption{Schematic diagram of the proposed metascreen for two-sided field modulation in contrast to other metasurfaces. The solid arrow indicates the modulated acoustic fields whereas the dashed arrow represents the non-modulated fields. The schematic diagram on the right illustrates a panel consisting of multiple two-sided acoustic metascreen unit cells, with each square unit representing a single cell.}\label{fig1}
\end{figure}
\\
In this paper, we introduce and demonstrate the functionality of a two-sided acoustic metascreen (TAM) aimed at extending acoustic wave manipulations to both reflection and transmission fields. Figure 1 presents a schematic diagram illustrating the unconventional properties of our proposed design. Unlike other acoustic metamaterials and metasurfaces are limited to modulating the phase profile of only the reflected or transmitted acoustic waves, the proposed TAM design enables simultaneous amplitude and phase control of both reflection and transmission waves. In this work, the full-space control capability of the reported structure is proved thoroughly via analytical derivation, numerical simulation, and experimental demonstration. \\
Motivated by the principles underlying lossy acoustic metamaterials\cite{zhu2018fine} and reconfigurable acoustic metascreen\cite{fan2023ultrabroadband} designs, our developed TAM achieves effective broadband wave modulations on both sides of the device simultaneously. Leveraging the decoupled tuning effect provided by the holey-structured lossy metamaterial, TAM facilitates independent control of both amplitude and phase of reflection and transmission waves across their full range ([0, 1] and [0, 2$\pi$]), spanning from 4000 Hz to 8000 Hz. Additionally, the incorporation of simple slit structures offers the benefits of significantly reducing fabrication costs and enabling efficient implementation. \\
To showcase its effectiveness, we first present simultaneous acoustic diffusion and focusing functionalities. In the next step, we further demonstrate its broadband characteristics by simultaneously steering and focusing acoustic waves at three distinct frequencies (5500 Hz, 6000 Hz, and 6500 Hz). Finally, through the implementation of the Iterative Angular Spectrum Approach (IASA) and the systematic design process of unit cells, we successfully demonstrated a two-sided acoustic hologram functionality (i.e., the hologram displays one alphabet letter on the reflection side and a different alphabet letter on the transmission side). Numerical simulations and experimental tests are conducted to confirm the validity of these intriguing results. This TAM design opens up an alternative way to design acoustic metamaterials that offer amplitude and phase modulations in both reflection and transmission domains. Additionally, parts of the result are demonstrated at three different frequencies, further underscoring the versatility of our proposed structure. Its remarkable flexibility in generating two acoustic fields individually holds significant potential for applications in communication, signal encryption, and imaging systems. 
\section{Results}\label{sec2}
\subsection{Unit cell design}
Figure 2(a) illustrates the operational concept of a TAM, which incorporates multiple unit cells designed in alignment with the principles of Snell's law. This array design enables the metascreen to attain diverse functionalities on both sides of the interface. Figure 2(b) shows a three-dimensional model of a unit cell, comprising a straight acoustic wave channel in a depth of $D=14.3$ mm. More detailed parameters can be found through the cross-section view of the unit cell, as shown in Fig. 2(c). At a distance of $h_1$ from the above inlet, the channel is decorated with a pair of protruding rectangular-shaped slabs with a thickness of $h_2$, the distance between these two slabs is set as $w$. Then, at a distance of $h_3$ from the slabs, there is a series of fifteen thin plates with a width of $w_2$. While the total height of the upper region, $h = h_1 + h_2 + h_3$, is set as constant at $h = 50$ mm, the phase of the reflected wave is determined by the positioning of the slabs, specifically, the value of $h_1$. In the meantime, the bottom thin plates remain in fixed positions but vary in their width $w_2$, which, in turn, controls the phase of the transmitted waves. It's worth noting that not only can the phase of waves be adjusted, but the amplitude of the reflected waves is also intricately connected to the structural parameter $w$. Considering the continuity of acoustic pressure across the interface, the amplitude of transmitted waves is further tuned. In general, this TAM design realizes the decoupled modulation of the phase and amplitude of both reflected and transmitted waves by individually adjusting three parameters $h_1, w_2$, and $w$. \\
Due to the significant acoustic impedance mismatch between the frame material and the background medium (air), the walls of the channels are considered acoustically rigid. Given that the TAM unit cell can be conceptualized as air channels with varying cross-sections, the acoustic impedance is primarily influenced by the ratio of cross-sections at interfaces and the length of each region. To evaluate the wave modulation capabilities of the unit cell, we utilized the transfer-matrix method, enabling the theoretical extraction of the phase and amplitude of both the reflected and transmitted waves.
In the steady state, acoustic waves traveling along the $z$-axis can be expressed as $p(z)=p_+\exp (-jk_0z) + p_-\exp (jk_0z)$, indicating two traveling waves: one with amplitude $p_+$ traveling in the positive $z$ direction and the other with amplitude $p_-$ traveling in the negative $z$ direction, where $k_0$ is the wave number of acoustic waves in air. For the TAM unit cell, we only consider the first order of acoustic waves, which means we ignore the echo of the acoustic wave from the following boundaries. At the $i$th interface, since there is no change in density across the interface, and conservation of mass requires that the volume velocity also be reserved, the incident, reflected, and transmitted pressure are connected as
\begin{equation}
p_t^{(i)} \exp \left(-j k_0 h_e\right) +p_r^{(i)} \exp \left(j k_0 h_e\right)=p_t^{(i+1)}+p_r^{(i+1)}, 
\end{equation}
\begin{equation}
\frac{p_t^{(i)} \exp \left(-j k_0 h_e\right) -p_r^{(i)} \exp \left(j k_0 h_e\right)}{S_i}=\frac{p_t^{(i+1)}-p_r^{(i+1)}}{S_{i+1}},
\end{equation}
where $h_e$ refers to the effective length of the upstream channel, $S_i$ is defined as the ratio of the width of the area-changing region to the width of the waveguide. Then, a transfer matrix at the $i$th interface $T_i$ can be used to relate the reflected ($p_r$) and transmitted ($p_t$) acoustic pressure at the upstream (subscript, $i$) and downstream (subscript, $i+1$) channel in the form of 
\begin{equation}
\left[\begin{array}{l}
p_t \\
p_r
\end{array}\right]_{(i+1)}=T_i\left[\begin{array}{l}
p_t \\
p_r
\end{array}\right]_{(i)}.
\end{equation}
Given that the TAM unit cell is composed of $i$ area-changing regions, the total transfer matrix $T_t$ to describe the acoustic behavior is expressed as
\begin{equation}
T_t=T_i T_{i-1} \cdots T_1,
\end{equation}
The detailed theoretical analysis of the TAM unit cell is given in Supplementary information.
\begin{figure}[!h]
\centering
\includegraphics[width=1\textwidth]{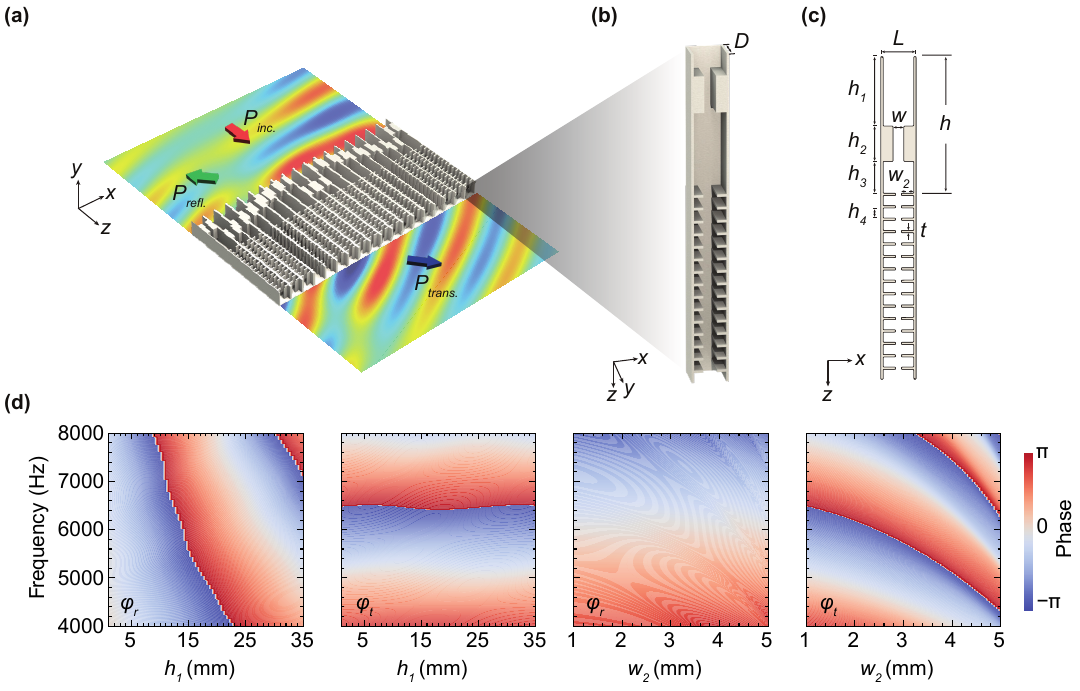}
\caption{Conceptual illustration of the simultaneous modulation on reflection and transmission wave. (a) Schematic diagram of an array of TAM unit cells. (b) A three-dimensional illustration of a unit cell. (c) Cross-section view of a unit cell and its geometrical parameters. The length of the unit cell $L=14.3$ mm, the thickness of the pair of slabs $h_2 = 14.3$ mm, the thickness of solid frames $t=1$ mm, the distance between adjacent thin plates $h_4 = 4$ mm. (d) Broadband simulated reflection and transmission phase responses to the parameters ${h_1}$ and ${w_2}$, respectively.}\label{fig2}
\end{figure}
Simulations are performed to verify the broadband wave manipulation capabilities of the TAM unit cell. The broadband and independent control of the phase of reflected and transmitted waves is achieved through two distinct geometric parameters, $h_1$ and $w_2$, as illustrated in Fig. 2(d). 
The first two plots in Fig. 2(d) demonstrate simulated phase of reflected ($\varphi_r$) and transmitted ($\varphi_t$) waves as a function of parameter $h_1$ across the frequency range of 4000 Hz to 8000 Hz, respectively. With $w_2$ set as 1 mm and all other parameters fixed, $h_1$ is adjusted in the range of 1 to 35 mm, resulting in a full coverage of 2$\pi$ phase change of the reflected waves in the frequency range of 4000 Hz to 8000 Hz. Notably, the phase of the transmitted waves remains constant during the adjustment of $h_1$ across the frequency range. Alternatively, when $h_1$ is set to 31 mm, the value of $w_2$ is varied within the range of 1 to 5 mm while keeping all other parameters fixed. As illustrated in the last two plots in Fig. 2(d), complete coverage of a 2$\pi$ phase shift of the transmitted waves is achieved, with little impact on the phase of the reflected waves. \\
According to the results in Fig. 2(d), the TAM unit cell is capable of independently modulating the phase of reflected and transmitted waves by adjusting two unrelated parameters, $h_1$ and $w_2$. Importantly, the proposed TAM unit cell can realize full coverage of a 2$\pi$ phase modulation, not only at a specific frequency but across broadband frequencies ranging from 4000 Hz to 8000 Hz. This extends beyond the capabilities of Janus metamaterials, which typically operate within a limited frequency range.
\begin{figure}[!h]
\centering
\includegraphics[width=1\textwidth]{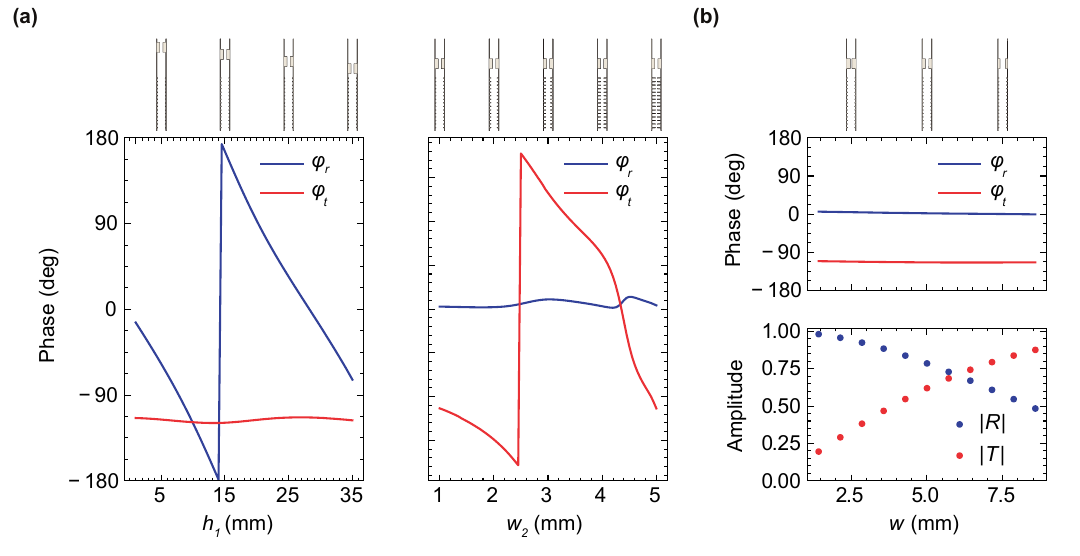}
\caption{At 6000 Hz, the phase and amplitude responses of the TAM unit cell correspond to various configurations: (a) Simulated phase as a function of $h_1$ when $w_2 = 1$ mm (left). Simulated phase as a function of $w_2$ when $h_1 = 28$ mm (right). (b) Simulated amplitude as a function of $w$ when $h_1$ and $w_2$ are fixed as constants.}\label{fig3}
\end{figure}
\\
To apply the capability of the TAM design to various functionalities, we select 6000 Hz as the operational frequency to design the TAM unit cell and arrays. In Figure 3, the top inserts correspond to the different configurations of the unit cell under various $h_1$, $w_2$, and $w$, respectively. Figure 3(a) illustrates the correlation of the phase of reflected and transmitted waves to $h_1$ and $w_2$ for the two cases of $w_2 = 1$ mm and $h_1 = 28$ mm, respectively. Within the adjustment of $h_1$ and $w_2$ and all other parameters fixed, the phase of reflection and transmission waves are tuned and realize the full coverage of 2$\pi$ of the phase map. \\
Furthermore, the TAM can realize the power allocation between reflection and transmission, enabling comprehensive control over acoustic waves in the full space. If we keep $h_1$ and $w_2$ and all other parameters as constants but modify the value of $w$, we can determine the allocation of acoustic energy to be reflected or transmitted, as shown in Fig. 3(b). As the value of $w$ approaches 0 mm, almost all acoustic energy is reflected since the opening of the channel is completely blocked. Conversely, as $w$ increases, more energy is transmitted instead of being reflected. Figure 3 demonstrates the decoupled modulation of amplitude and phase of both reflection and transmission waves within the full range of [0, 1] and [0, 2$\pi$]. By adjusting parameters $h_1$, $w_2$, and $w$, the TAM unit cell can access all possible combinations of amplitude and phase ($A$, $\varphi$), allowing it to easily deliver versatile acoustic waves upon request. Such an extraordinary feature enables a complete manipulation of sound in the full space.
\subsection{Independently modulate reflection and transmission fields}
Following the systematic design of the unit cell, we aim to demonstrate the complete control of acoustic waves in full space by designing and arranging the TAM unit cells according to the generalized Snell's law. Consider plane waves incident on a linear boundary, where the bending direction of the waves is related to the phase gradient in the form of\cite{yu2011light,ni2012broadband}
\begin{equation}
\sin (\theta_t) n_t-\sin (\theta_i) n_i=\lambda_0 \nabla \Phi / 2 \pi,\label{eq3}
\end{equation}
where $\nabla \Phi=d \Phi / d z$ indicates the gradient of the phase discontinuity along the boundary. Therefore, by designing and engineering a phase gradient along an interface, we can control the waves’ bending direction to deliver versatile functionalities. Integrated with the capability of the TAM unit cell to modulate the phase of reflection and transmission waves, the wave pattern on both sides of the unit cells can be controlled independently.
\begin{figure}[!h]
\centering
\includegraphics[width=1\textwidth]{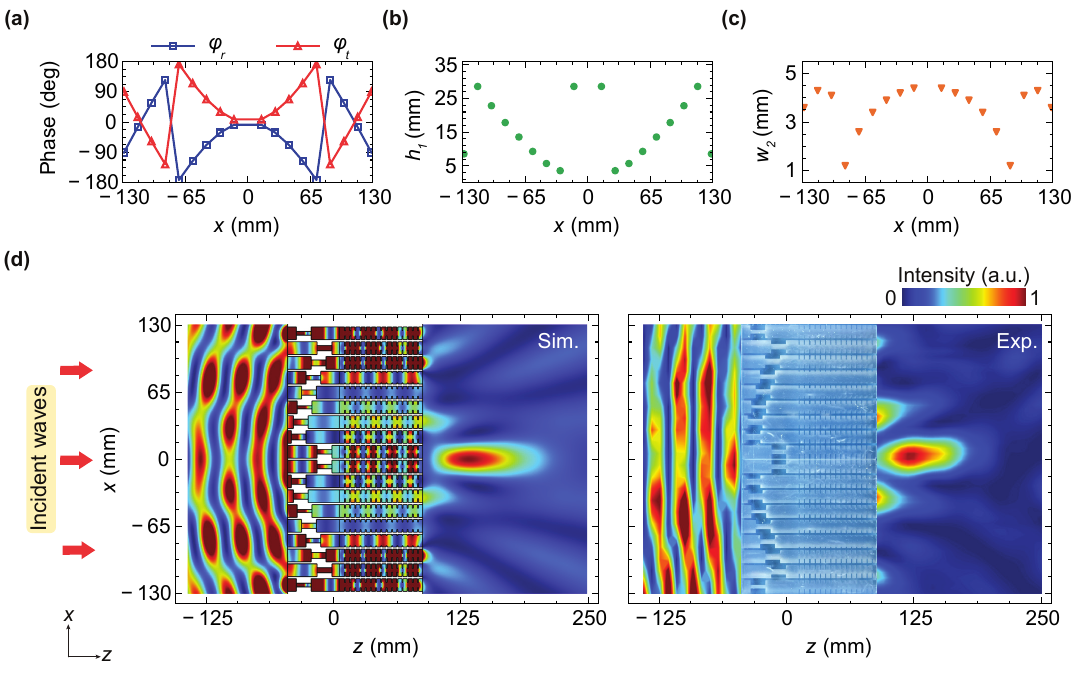}
\caption{The designed TAM array enables acoustic diffusion on the reflection side and acoustic focusing on the transmission side. (a) The phase profiles for the acoustic diffusion and focusing are calculated according to the generalized Snell's law. (b) The corresponding $h_1$ values for the acoustic diffusion function vary with the $x$-coordinates. (c) The corresponding $w_2$ values for the acoustic focusing function vary with the $x$-coordinates. (d) Simulated (left) and measured (right) acoustic intensity at 6000 Hz with plane waves incident from the left side. The insert is a photograph of our 3D-printed TAM array.}\label{fig4}
\end{figure}
\\
We first examine the case of achieving acoustic diffusion and focusing simultaneously within a TAM array at 6000 Hz, as illustrated in Fig. 4. On the reflection side, the normal incident plane waves are manipulated to free diffusion, while the transmitted waves are focused at a focal point. In simplicity, acoustic diffusion can be regarded as the backward wave path of acoustic focusing, resulting in opposite phase values of the reflection and transmission waves for each unit cell. The phase profile of the TAM array for acoustic diffusion ($\varphi_r$) and focusing ($\varphi_t$) are obtained by
\begin{equation}
\varphi_r=k_0\left(\sqrt{z_0^2+x_0^2}-\sqrt{\left(z-z_0\right)^2+\left(x-x_0\right)^2}\right),\label{eq4}
\end{equation}
\begin{equation}
\varphi_t=k_0\left(\sqrt{\left(z-z_0\right)^2+\left(x-x_0\right)^2}-\sqrt{z_0^2+x_0^2}\right),\label{eq5}
\end{equation}
where $(z_0, x_0)$ is the coordinate of the desired focal point. When the TAM array is symmetric with respect to the $x$-axis, and in the case of acoustic waves being focused at (160 mm, 0), the phase profiles of the TAM array along the $x$-axis are presented in Fig. 4(a). The corresponding $h_1$ and $w_2$ values for each unit cell are shown in Fig. 4(b) and (c), respectively. Finally, the reflection and transmission intensity fields for this TAM array are numerically simulated and experimentally measured, as shown in Fig. 4(d). The similarities between these two results confirm the capability of the proposed TAM to individually modulate acoustic waves on both the reflection and transmission sides. Such a feature of encoding two functionalities into a single array will increase freedom for signal encryption and information processing.
\subsection{Broadband two-sided wave manipulations}
One key advantage of the proposed TAM design compared to other existing structures is its ability to manipulate waves on both sides, not only at one specific frequency but in a broadband manner.
\begin{figure}[!h]
\centering
\includegraphics[width=1\textwidth]{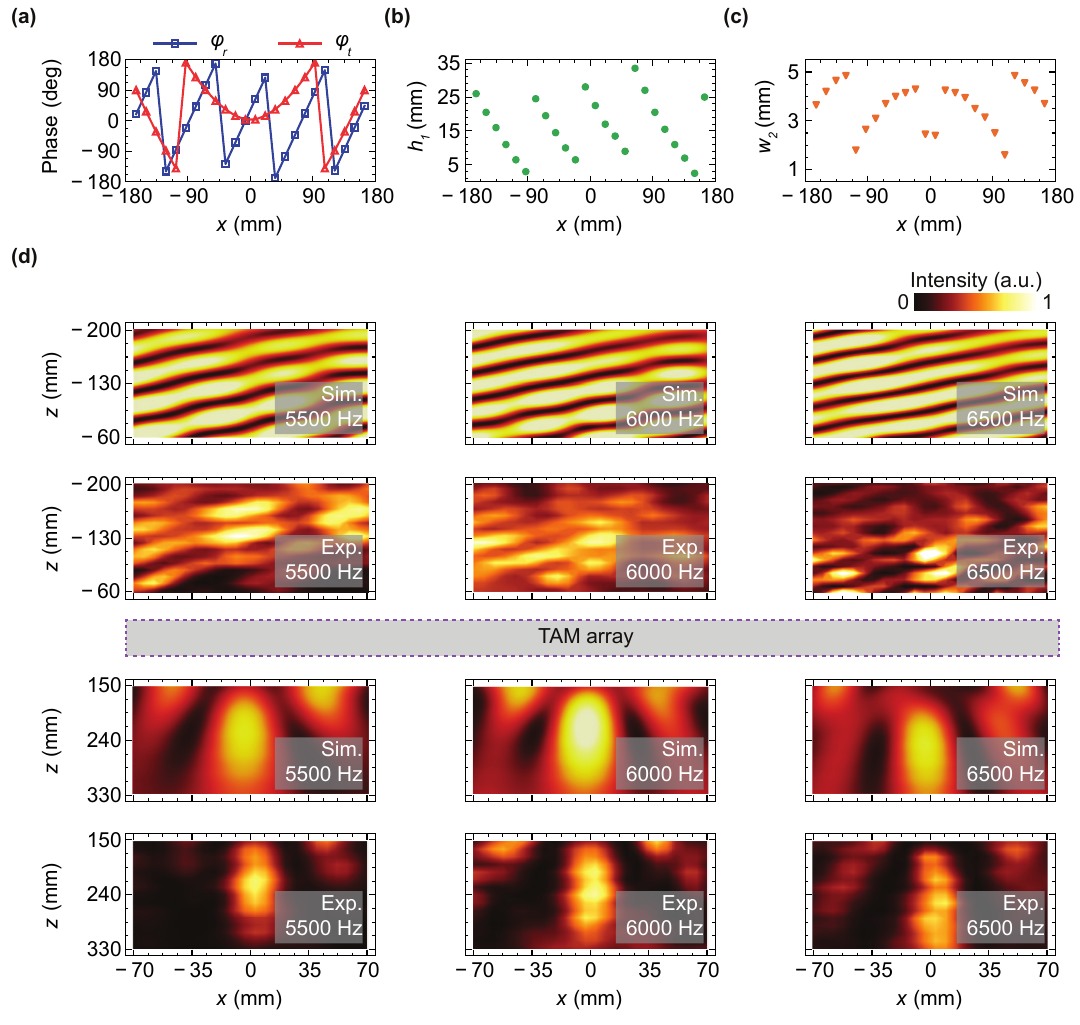}
\caption{The broadband TAM array design enables acoustic reflection for the reflection side and acoustic focusing for the transmission side across a broadband frequency range. (a) The phase profiles for the acoustic reflection and focusing are calculated according to the generalized Snell's law. (b) The corresponding $h_1$ values for the acoustic reflection function vary with the $x$-coordinates. (c) The corresponding $w_2$ values for the acoustic focusing function vary with the $x$-coordinates. (d) Simulated and measured acoustic intensity at 5500, 6000, and 6500 Hz with plane waves incident from the top side.}\label{fig5}
\end{figure}
\\
In the second example, another TAM array is designed to realize acoustic reflection at an angle of 45$^{\circ}$ on the reflection side and focus at coordinate (250 mm, 0) on the transmission side simultaneously at three selected frequencies, 5500, 6000, and 6500 Hz.
In this demonstration, 24 unit cells were designed to form the new TAM array to ensure the good demonstration quality of the acoustic reflection and focusing.
Following the same procedure, the phase profiles for the acoustic reflection ($\varphi_r$) and focusing ($\varphi_t$) at 6000 Hz are calculated at first, as depicted in Fig. 5(a). The phase profile for the acoustic reflection is determined via
\begin{equation}
\varphi_r=k_0 z \sin \theta_r,\label{eq6}
\end{equation}
where $\theta_r$ represents the reflection angle. Then, geometrical parameters $h_1$ and $w_2$ are selected based on data points in Fig. 2(d) and shown as discrete points in Fig. 5(b-c). Simulations and experiments are performed at 5500, 6000, and 6500 Hz to demonstrate the broadband two-sided wave manipulation capability of the designed TAM array, as shown in Fig. 5(d). Although the phase profiles of the TAM array were only optimized for reflection and focusing functionalities at 6000 Hz, the reflection of plane wave at 45 degrees and convergence of acoustic energy at a focal point can still be easily observed at the other two selected frequencies. These results further validate the capability of the proposed metascreen unit design to realize full coverage of 2$\pi$ phase shifts at both sides in a broadband manner. 
\subsection{Broadband two-sided acoustic holograms}
The above sections successfully demonstrate the capabilities of the proposed design to manipulate acoustic waves on both sides in a broadband frequency range. To further showcase the ultimate acoustic wave manipulations, the proposed TAM design was applied to build a three-dimensional (3D) panel consisting of 625 (25$\times$25) TAM unit cells. A function of broadband two-sided acoustic holograms is numerically simulated and experimentally demonstrated. Notably, the two-sided acoustic holograms function shown here is different from the reported Janus holograms. Unlike Janus holograms, which require changing the direction of incident waves, the design proposed here can realize two distinct holograms on the reflection and transmission sides simultaneously.
\begin{figure}[!h]
\centering
\includegraphics[width=1\textwidth]{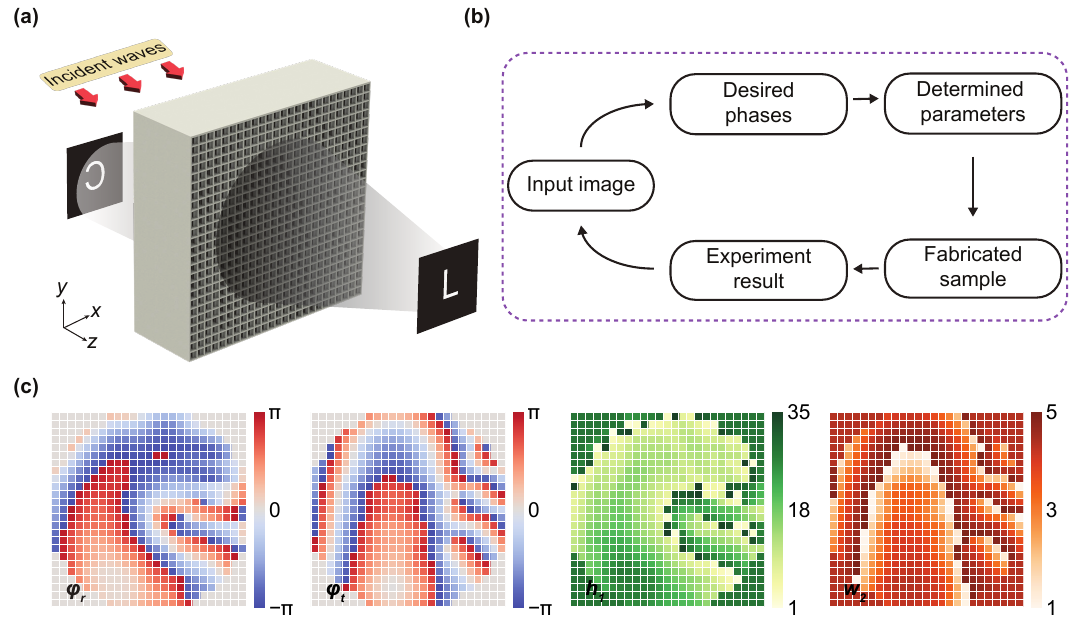}
\caption{The designed TAM panel for acoustic holograms. (a) Schematic diagram of the TAM panel, generating two different letters on the reflection and transmission side. (b) Workflow of the TAM-based acoustic holograms. (c) The determined phase profiles and the corresponding geometrical parameter maps.}\label{fig6}
\end{figure}
\\
Compared to wave modulations in two dimensions, the acoustic hologram is excellent for demonstrating the flexibility and encoding functionality of a structure. The working scenario of the proposed TAM-based acoustic holograms is shown in Fig. 6(a). With plane waves incident from the rear of the TAM panel, two different holograms are generated simultaneously on the reflection and transmission sides. The design procedure is shown in Fig. 6(b). \\
To generate the two distinct acoustic holograms, (e.g., the binary amplitude of two letters 'C', and 'L'), the IASA was employed to determine the phase profiles of the reflection and transmission sides of the TAM panel\cite{liu1997propagation,melde2016holograms,xu2023sound}. Considering each TAM unit cell as a hologram pixel and the hologram plane located at $z=0$ plane, the acoustic pressure wave is expressed as
\begin{equation}
p(x, y, 0)=\hat{p}(x, y, 0) e^{j \Delta \varphi(x, y, 0),}
\end{equation}
where $\hat{p}(x, y, 0)$ represents the pressure map and $\Delta \varphi(x, y, 0)$ is the phase map. The angular spectrum of this wave at plane $z=0$ is obtained by the Fourier transform
\begin{equation}
P\left(k_x, k_y, 0\right)=\iint_{-\infty}^{+\infty} p(x, y, 0) e^{-j\left(k_x x+k_y y\right)} \mathrm{d} x \mathrm{~d} y.
\end{equation}
The propagator function is
\begin{equation}
H\left(k_x, k_y, z\right)=e^{j z \sqrt{k^2-k_x^2-k_y^2}}.
\end{equation}
The angular spectrum at plane $z$ can be calculated by multiplying the angular spectrum with the propagator function
\begin{equation}
P\left(k_x, k_y, z\right)=P\left(k_x, k_y 0\right) H\left(k_x, k_y, z\right).
\end{equation}
Finally, the real-space pressure field in any plane z can be obtained via the inverse Fourier transform
\begin{equation}
p(x, y, z)=\frac{1}{4 \pi^2} \iint_{-\infty}^{+\infty} P\left(k_x, k_y, z\right) \mathrm{d} k_x \mathrm{~d} k_y.
\end{equation}
The IASA involves transforming the pressure field from the hologram plane into the angular spectrum, which is then propagated to the target plane (70 mm away from the TAM panel, $z=-120$ and 150 mm, respectively). Subsequently, the angular spectrum of the target plane is transformed into the wave function, and the quality of the resulting pressure field is evaluated until the maximum number of iterations is reached. Based on the pressure field obtained at the hologram plane, the geometrical parameters are determined using detailed data points available in Fig. 2(d). For each unit cell of the TAM panel, two sets of parameters, $h_1$ and $w_2$, are determined to achieve the best match of phase profiles for the reflected and transmitted waves. The resulting phase profiles and parameter maps are presented in Fig. 6(c) as a result of this process. Following the determination of geometrical parameters, the TAM panel is fabricated using a commercialized 3D printer.
\begin{figure}[!h]
\centering
\includegraphics[width=1\textwidth]{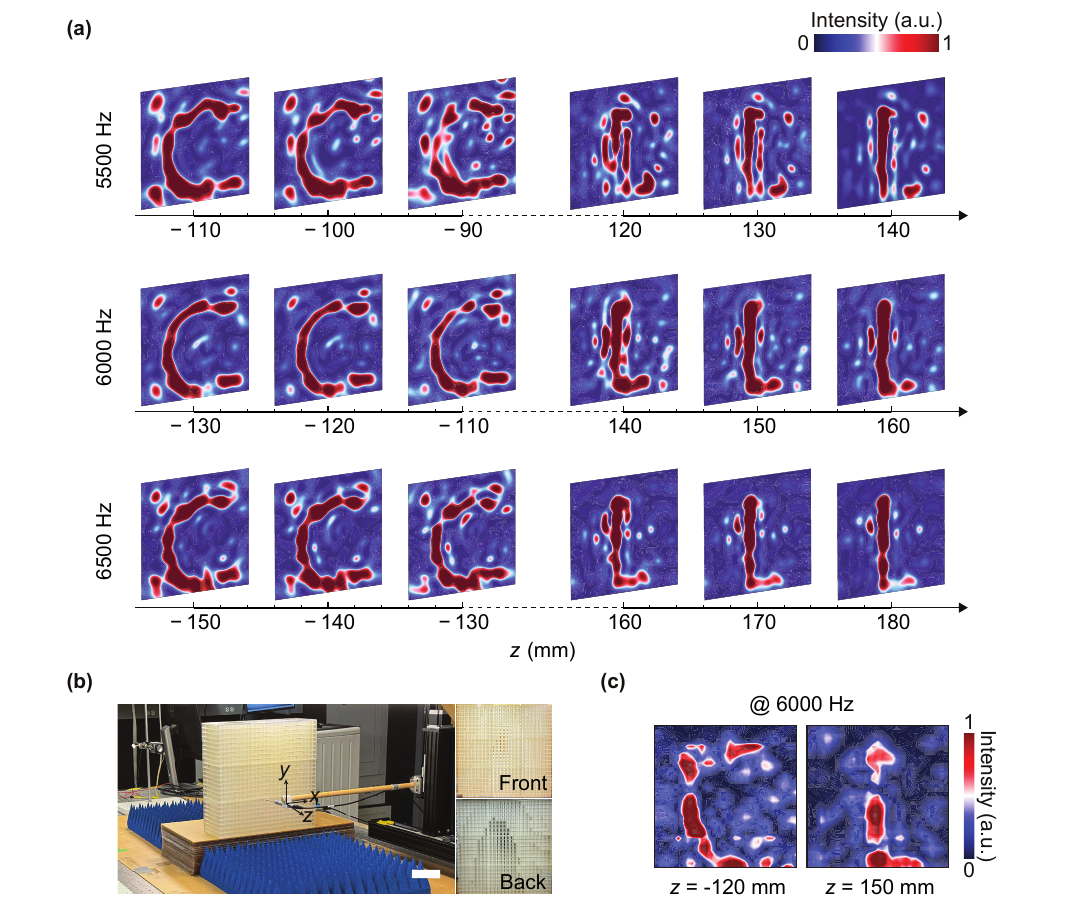}
\caption{(a)Simulated holographic images of the letter "C" on the reflection side and the letter "L" on the transmission side are observed at different observation planes for three selected frequencies: 5500 Hz, 6000 Hz, and 6500 Hz. The arrow denotes the direction of wave propagation along the $z$-direction. (b) Photographs of the experimental setup and the 3D-printed TAM panel. Scale bar: 10 cm. (c) Experimentally Measured holograms at 6000 Hz.}\label{fig7}
\end{figure}
\\
Figure 7(a) presents simulated acoustic intensity distributions on the reflection and transmission side of the TAM panel for three distinct frequencies (5500, 6000, and 6500 Hz) of incidence. The arrows in Fig. 7(a), indicate the direction of propagation waves and show the numerically calculated intensity field at different $z$ planes. It is worth noting that the IASA method was conducted with the frequency set at 6000 Hz, indicating that for the other two frequencies, the phase profiles are not optimized. Due to the wavelength variation with operational frequency changes, the letters "C" and "L" appear in different planes at different frequencies. Nevertheless, they remain distinctly visible in all three cases, highlighting the TAM panel's full-space modulation capability. \\
In the experiment, we measure the spatial distributions of sound intensity across two $z$ planes before and after the TAM panel using a 1/8-inch pressure-field microphone (Bruel \& Kjaer type-4138C). The microphone is affixed to a 3D stepping motor, enabling precise scanning of the target planes point by point. The photograph of the experimental setup is shown in Fig. 7(b), with the inserts referring to the front ($z=-50$ mm) and back ($z=80$ mm) surfaces of the 3D-printed TAM panel. Lastly, the measured acoustic holograms are illustrated in Fig. 7(c), with the letter "C" appearing at the plane $z=-120$ mm and the letter "L" appearing at the plane $z=150$ mm at 6000 Hz. A good agreement was observed when compared with the results in Fig.7(a).
\subsection{Discussion}
In summary, we have presented a design featuring air channels with varying cross-sections to achieve full-space modulation of acoustic waves across a broad frequency range. The proposed two-sided acoustic metascreen, termed TAM, demonstrates significant capabilities in wave manipulation, as validated through numerical simulations and experimental demonstrations showcasing two-sided acoustic wave functionalities, including diffusion, reflection, and focusing. Moreover, leveraging the IASA method and the capacity of decoupled reflection and transmission control, we have demonstrated the two-sided acoustic holograms. We envision that our findings regarding independent control of reflection and transmission fields will spark interest in various applications relying on acoustic waves, such as particle manipulation, architectural acoustics, and stereo sound-field reconstruction, among others.
\section{Methods}
\subsection{Numerical simulations}
Numerical simulations were performed using the finite element software COMSOL Multiphysics 6.0, “Pressure Acoustics, Frequency Domain (acpr)” module. In all simulations, the background medium is set as air ($\rho_0=1.22 \mathrm{~kg} / \mathrm{m}^3$, $c_0=343 \mathrm{~m} / \mathrm{s}$). The outer frames of solids are set as sound-hard boundaries. In Fig. 2-7, frequency domain studies are deployed to determine the phase and amplitude of acoustic reflection transmission, as well as the intensity fields. Background pressure fields are set with an amplitude of 1 Pa. Perfectly matched layers (PML) are adopted to reduce unwanted reflections at surrounding boundaries.
\subsection{Experimental measurements}
The TAM units were fabricated by a commercial stereolithography-based 3D printer (Form 3+). The TAM arrays shown in Fig. 4-5 were assembled with fabricated TAM units and placed in a planar waveguide. The top and bottom sides of the waveguide were covered with acrylic boards, while the sides of the waveguide were filled with absorbing foam to reduce unnecessary acoustic reflection at the boundaries. A speaker was placed enough far away from the TAM array to generate plane waves. The reflected and transmitted acoustic pressure fields are acquired with two microphones (Bruel\&Kjaer type 4138C microphone), with one fixed close to the speaker to record as a reference signal, while the other was attached to a two-dimensional motorized linear stage (VELMEX BiSlide) on the transmission side. Point-to-point scanning was performed with a step of 1 cm in the area of interest. The measured data were recorded and analyzed by the Bruel\&Kjaer 3160-A-042 module. The intensity fields in Fig.7 are measured at the image plane with an area of 31 × 35 \si{cm^2}.
\section*{Data availability}
All relevant data that support the findings of this study are available from the corresponding authors upon reasonable request.
\bibliography{sn-bibliography}
\newpage
\section*{Acknowledgments}
The authors would like to thank Boston University Photonics Center and the Rajen Kilachand Fund for Integrated Life Science and Engineering for funding and technical support.
\section*{Author contributions}
A.C. and X.Z. conceived the idea and interpreted the results. A.C. designed and performed the theoretical and numerical calculations. A.C. under guidance of X.Z. designed and performed the experiment. A.C. and X.Z. contributed to the preparation and writing of the manuscript, and X.Z. planned, coordinated, and supervised the project.
\section*{Competing interests}
The authors declare no competing interests.
\section*{Supplenmentary information}
The online version contains supplementary material available at this address:
\section*{Materials \& Correspondence}
Correspondence and requests for materials should be addressed to Ao Chen or Xin Zhang.

\end{document}